\documentclass[useAMS,usenatbib]{mn2e}

\usepackage{epsf,rotating}
\usepackage{amsmath}
\usepackage{subfigure}

\newcommand{\fgas}{f_{\rm gas}}
\newcommand{\mgas}{M_{\rm gas}}
\newcommand{\tdep}{t_{\rm dep}}

\title[An Analytic Model for Galaxy Evolution]{An Analytic Model for the
Evolution of the Stellar, Gas, and Metal Content of Galaxies}

\author[Dav\'e, Finlator, \& Oppenheimer]{
\parbox[t]{\textwidth}{\vspace{-1cm}Romeel Dav\'e$^1$, Kristian Finlator$^2$, Benjamin D. Oppenheimer$^3$}
\\\\$^1$ Astronomy Department, University of Arizona, Tucson, AZ 85721, USA
\\$^2$ Hubble Fellow; Physics Department, University of California, Santa Barbara, CA 93106, USA
\\$^3$ Veni Fellow; Leiden Observatory, Leiden University, PO Box 9513, 2300 RA Leiden, Netherlands
}

\begin{document}

\maketitle

 \begin{abstract}
We present an analytic formalism that describes the evolution of
the stellar, gas, and metal content of galaxies.  It is based on
the idea, inspired by hydrodynamic simulations, that galaxies live
in a slowly-evolving equilibrium between inflow, outflow, and star
formation.  We argue that this formalism broadly captures the
behavior of galaxy properties evolving in simulations.  The resulting
equilibrium equations for the star formation rate, gas fraction,
and metallicity depend on three key free parameters that represent
ejective feedback, preventive feedback, and re-accretion of ejected
material.  We schematically describe how these parameters are
constrained by models and observations.  Galaxies perturbed off the
equilibrium relations owing to inflow stochasticity tend to be
driven back towards equilibrium, such that deviations in star
formation rate at a given mass are correlated with gas fraction and
anti-correlated with metallicity.  After an early gas accumulation
epoch, quiescently star-forming galaxies are expected to be in
equilibrium over most of cosmic time.  The equilibrium model provides
a simple intuitive framework for understanding the cosmic evolution
of galaxy properties, and centrally features the cycle of baryons
between galaxies and surrounding gas as the driver of galaxy growth.
\end{abstract}

\section{Introduction} 

Galaxy formation involves a wide range of diverse physical processes
operating on stellar to cosmological scales, including the hierarchical
growth of structure, star formation, black hole accretion, and a
plethora of poorly-understood feedback processes that strongly
modulate galaxy growth.  Given this complexity, it is surprising
that galaxies display simple and tight scaling relations between
many of their key constituents.  These include well-established
relations between the bulge velocity dispersion and central black
hole mass~\citep[e.g.][]{gul09,gra11}, circular velocity and luminosity~\citep{tul77}, star formation rate and stellar mass~\citep[e.g.][and
references therein]{dav08,gon10}, and metallicity and stellar
mass~\citep[e.g.][]{tre04,erb06}.  Each has low scatter and evolves
roughly independently of mass.  The simplicity of these relations
hints at an underlying uniformity in galaxy evolution that is not
immediately evident from the complexity of current hierarchical
galaxy formation models.

In the longstanding canonical scenario for galaxy formation, galaxies
form as angular momentum-conserving disks cooling from hot gas bound
within dark matter halos, and these disks subsequently merge to
form larger and earlier-type galaxies~\citep{ree77,whi78,whi91,mo98}.
This scenario is well-situated within currently favored hierarchical
cosmologies, and analytic models based on it have been quite
successful at reproducing many observed galaxy properties~\citep[see
review by][]{ben10}.  However, the present generation of such models
(often called ``semi-analytic" models) are typically enormously
complex, with a host of free parameters describing various interrelated
physical phenomena.  Numerical simulations that explicitly track
gas dynamical processes enable a more ab initio calculation, but
still require many ``sub-grid" parameters for key physical processes
and are in practice limited by dynamic range and numerical
uncertainties.  In either case, the complexity of such models makes
it difficult to extract simple physical intuition for what drives
the evolution of basic galaxy properties.

In this paper, we present an analytic framework for understanding
the evolution of the stellar, gas, and metal content of galaxies.
This framework is based on intuition gained from hydrodynamic
simulations of galaxy formation.  In such models, galaxies are fed
primarily by cold ($\sim 10^4$~K) streams connecting to filamentary
large-scale structure~\citep{ker05,dek09}, outflows are strong and
ubiquitous~\citep{spr03b,opp08}, and outflowing material commonly
returns to galaxies~\citep[``wind recycling";][]{opp10}.  Hence in this
framework, galaxy evolution is governed by the cycle of baryons exchanging
matter and energy between galaxies and surrounding intergalactic gas.

Our framework is an attempt to distill the insights gained from
such hydrodynamic simulations into an analytic formalism that both
describes the results of simulations and provides intuition into the key
physical drivers.  It is based primarily on the formalism presented in
\citet{fin08}, with key extensions, and shares features with various
recent works~\citep[e.g.][]{ras06,bou10,dut10,kru11}, indicating a
groundswell towards this ``baryon cycling"\footnote{A term coined in
the Astro2010 Decadal Survey Report: New Worlds, New Horizons} view
of galaxy formation.  We demonstrate that the star formation rate, gas
content, and metallicity of galaxies can be described by simple equations
that depend on three parameters that are directly related to inflows,
outflows, and wind recycling.  These parameters are poorly constrained in
both value and functional form, and hence the number of free parameters
in this model may be much greater than three, pending observations that
can better constrain them.  Together, these parameters quantify
the impact of baryon cycling on galaxy growth.  We give examples of
how these equations lead to straightforward intuitive explanations,
often differing from traditional ones, for the results seen in recent
observations and models of galaxy evolution.

This paper begins in \S\ref{sec:basics} by describing the basis for
our analytic framework, namely the equilibrium condition, and
discusses the physical constraints on its various terms that
ultimately govern stellar growth.  In \S\ref{sec:mgr} we present
an expression for gas fractions and explore some implications.
\S\ref{sec:mzr} discusses what governs galaxy metallicities, and
relates this to wind recycling.  \S\ref{sec:equil} gives some brief
examples of how these equilibrium relations yield straightforward
intuition into what governs basic galaxy properties.  \S\ref{sec:sample}
discusses a preliminary implementation of an equilibrium model, and
explores some parameter variations.  \S\ref{sec:scatter} considers
what happens when galaxies depart from equilibrium owing both to
stochastic fluctuations and more permanent departures.  \S\ref{sec:zeq}
discusses when galaxies first attain equilibrium in the early
universe.  Finally, we summarize and discuss broader implications
of our framework in \S\ref{sec:summary}.

\section{The Equilibrium Condition}\label{sec:basics}

Star-forming galaxies in hydrodynamic simulations are usually seen
to lie near the {\it equilibrium condition}~\citep[see e.g. Figure~13
of][and \citealt{dut10,bou10}]{fin08}:
\begin{equation}\label{eqn:equil}
\dot{M}_{\rm in} = \dot{M}_{\rm out} + \dot{M}_{\rm *},
\end{equation}
where the terms are the mass inflow rate, mass outflow rate, and
star formation rate (SFR), respectively.  Inflow and outflow refer
to gas motion in and out of the galaxy's star-forming region, i.e.
the interstellar medium (ISM).  Star-forming galaxies fluctuate
around this relation but are generally driven back to it on short
timescales, as we discuss in \S\ref{sec:scatter}; this ``self-regulating"
behavior is why we dub this model the {\it equilibrium model.}

The equilibrium condition is close to an expression for mass conservation,
except that it importantly does not contain a term describing a
gas reservoir.  A key ansatz of this formalism is that the rate of
change in the gas reservoir is small compared to the other terms
in Equation~\ref{eqn:equil}.  The motivation for this ansatz is that
in \citet{fin08}, we found this to be explicitly true in hydrodynamic
simulations of galaxy formation.  We note that this scenario has also been
referred to as a ``reservoir" or ``bath tub" model~\citep{bou10,kru11}.

Defining the mass loading factor $\eta\equiv \dot{M}_{\rm
out}/\dot{M}_{\rm *}$, we can rewrite the equilibrium condition as
\begin{equation}\label{eqn:sfr}
{\rm SFR} = \dot{M}_{\rm in}/(1+\eta).
\end{equation}
Hence in this scenario, 
a galaxy's star formation history over cosmic timescales is determined
by the evolution of $\dot{M}_{\rm in}$ and $\eta$.

Let us consider inflow first.  $\dot{M}_{\rm in}$ can be broadly
separated into three terms:
\begin{itemize}
\item $\dot{M}_{\rm grav}=$ Baryonic inflow into galaxy's halo, which is primarily
set by the assumed cosmology.  Here we employ the form forwarded by~\citet{dek09}:
\begin{equation}\label{eqn:Min}
\frac{\dot{M}_{\rm grav}}{M_{\rm halo}}
= 0.47 f_b \Bigl(\frac{M_{\rm halo}}{10^{12} M_\odot}\Bigr)^{0.15} \Bigl(\frac{1+z}{3}\Bigr)^{2.25}\;{\rm Gyr}^{-1}.
\end{equation}
\citet{fak10} presented a different parameterization of $25.3 M_{\rm
halo}^{0.1}(1+1.65z)\sqrt{\Omega_m(1+z)^3+\Omega_\Lambda}\;
M_\odot/$yr, while \citet{fau11} found $33.6 M_{\rm
halo}^{0.06}(1+0.91z)\sqrt{\Omega_m(1+z)^3+\Omega_\Lambda}\;
M_\odot/$yr; these yield similar results over most of cosmic time.

\item $\dot{M}_{\rm prev}=$ The amount of the gas entering the halo
that is {\it prevented} from reaching the ISM (to be subtracted
from the halo infall).  This is material that ends up in the gaseous
halo of the galaxy, and can also be regarded as the rate of growth
of halo gas.  We characterize this by defining a {\it preventive
feedback parameter}
\begin{equation}\label{eqn:zeta}
\zeta\equiv 1-\dot{M}_{\rm prev}/\dot{M}_{\rm grav}.
\end{equation}

\item $\dot{M}_{\rm recyc}=$ Gas infalling that has previously been
ejected in outflows, along with gas returned to the ISM via stellar
evolution.  This provides an extra component in addition to the
baryons associated with the gravitational infall of dark matter
(i.e. $\dot{M}_{\rm grav}$).

\end{itemize}
Expressing $\dot{M}_{\rm in}$ in these terms, we obtain 
\begin{equation}\label{eqn:Minsplit}
\dot{M}_{\rm in} = \dot{M}_{\rm grav}-\dot{M}_{\rm prev}+\dot{M}_{\rm recyc} = \zeta \dot{M}_{\rm grav} + \dot{M}_{\rm recyc}
\end{equation}

Halo infall ($\dot{M}_{\rm grav}$) is driven by gravity, hence is
mostly independent of feedback processes~\citep{vdv11,fau11} and
is determined primarily by cosmology and the halo merger
rate~\citep[e.g.][]{nei06,nei08,mcb09,genel10}.  In contrast,
$\dot{M}_{\rm prev}$ and $\dot{M}_{\rm recyc}$ are direct consequences
of feedback processes.  We will discuss $\dot{M}_{\rm recyc}$ further
in \S\ref{sec:mzr}, when we will relate it to the metallicity of
the infalling gas.  This leaves the preventive feedback parameter
$\zeta$, which we now consider.

There are a number of sources of preventive feedback, each with its own
dependence on halo mass and redshift, the products of which comprise the
total $\zeta$:

\begin{figure}
\vskip -0.4in
\setlength{\epsfxsize}{0.60\textwidth}
\centerline{\epsfbox{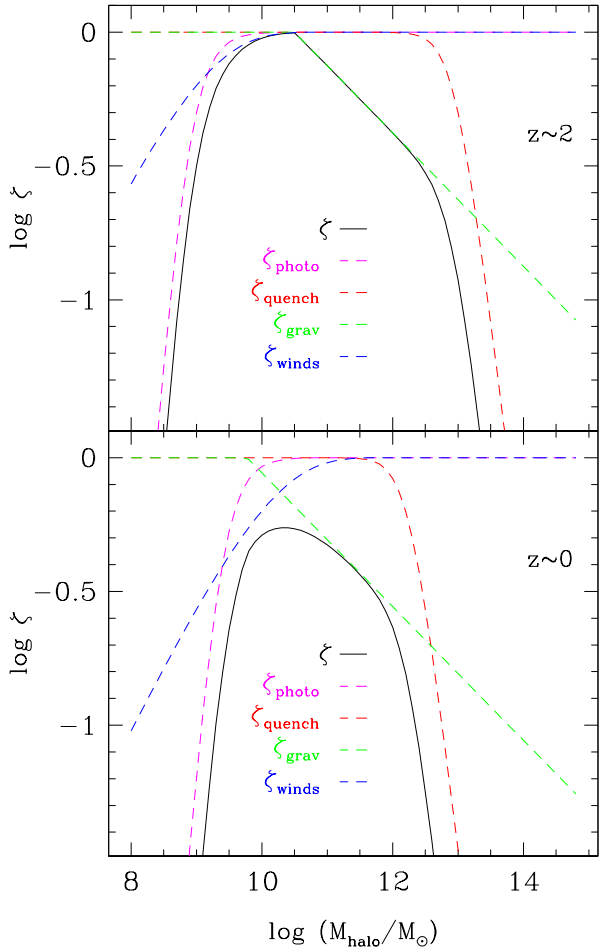}}
\vskip -0.2in
\caption{Schematic illustration of the preventive feedback parameter
$\zeta(M_{\rm halo})$.  $\zeta_{\rm photo}$ (magenta) is parameterized
here as $(1+\frac{1}{3}(M/M_\gamma)^{-2})^{-1.5}$~\citep[following
eq.~1 of][]{oka08}, and $\zeta_{\rm quench}$ (red) is analogously
parameterized.  $\zeta_{\rm grav}$ (green) is taken from
Equation~\ref{eqn:zetagrav}.  $\zeta_{\rm winds}$ (blue) is an
arbitrary function depicting an increasing effect to smaller masses
from wind heating.  {\it Top panel:} We choose $M_{\gamma}=10^9
M_\odot$ and $M_{\rm q}=10^{13} M_\odot$, perhaps appropriate
for $z=2$.  {\it Bottom panel:}  We choose $M_{\gamma}=3\times 10^9
M_\odot$, $M_{\rm q}=10^{12.3} M_\odot$, and a stronger $\zeta_{\rm winds}$,
more appropriate for $z=0$.  These values and their 
parameterizations are fairly
arbitrary, because the terms are poorly constrained; 
this plot is purely intended to illustrate general trends.
The total $\zeta=\zeta_{\rm photo}\zeta_{\rm quench}\zeta_{\rm grav}\zeta_{\rm winds}$ is shown as the black line, showing that vigorous
star formation can only proceed within an intermediate range of
halo masses where $\zeta$ approaches unity.
}
\label{fig:zeta}
\end{figure}

\begin{figure}
\vskip -0.7in
\setlength{\epsfxsize}{0.55\textwidth}
\centerline{\epsfbox{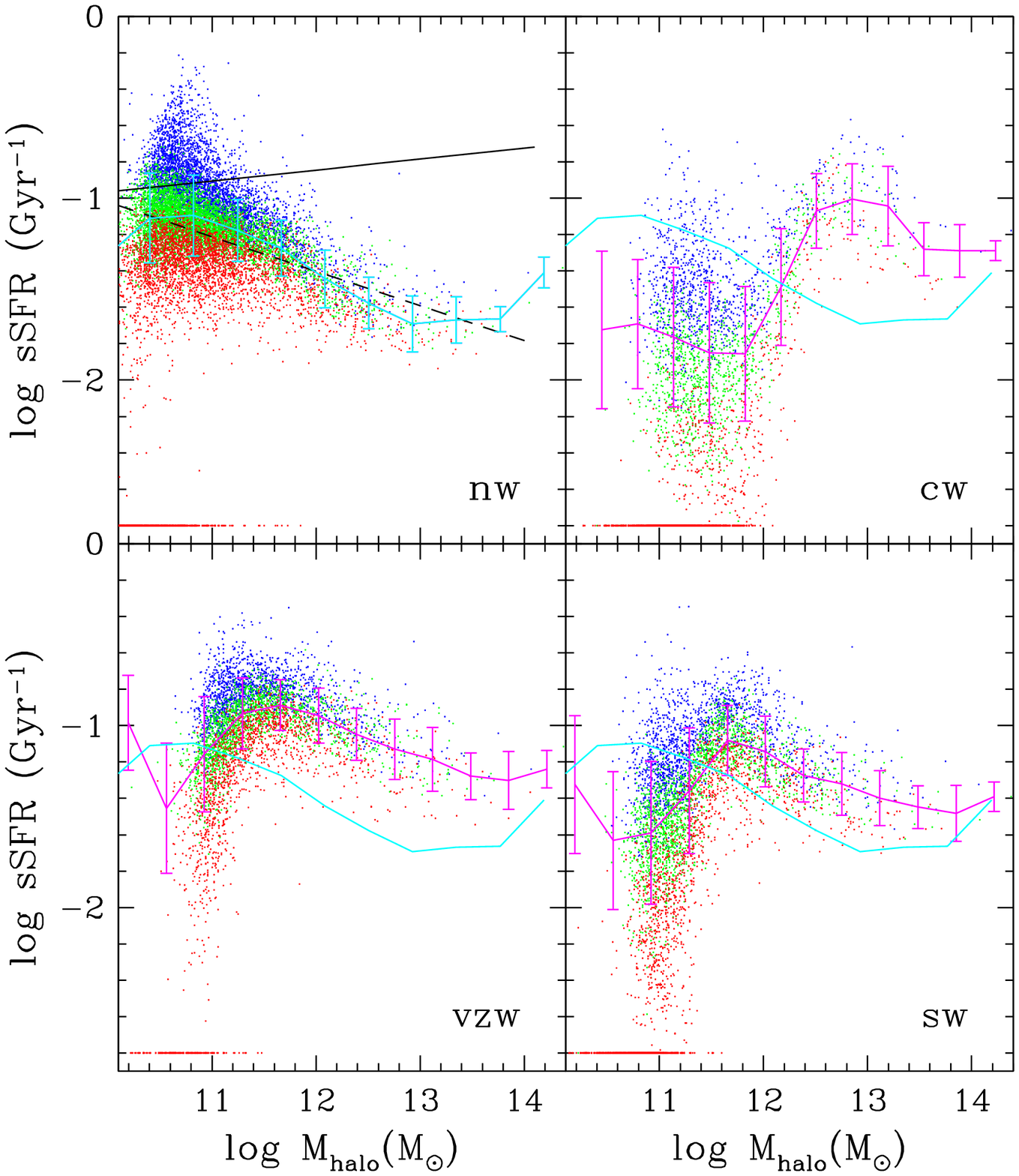}}
\vskip -0.6in
\caption{Specific star formation rate (sSFR$\equiv$SFR/$M_*$) versus
halo mass $M_{\rm halo}$ at $z=0$ in 48~Mpc/h cosmological hydrodynamic
simulations~\citep[see][for details]{dav11}.  Each point represents
a galaxy, and at a given mass these are subdivided into high $\fgas$
(blue; $>0.5\sigma$ above the median), low $\fgas$ (red; $<-0.5\sigma$
below the median), and intermediate (green).  Upper left panel shows the
case with no outflows (nw).  Solid black line shows the halo mass scaling
of $\dot{M}_{\rm grav}$ from \citet{fak10}, while dashed line shows the
effect of gravitational preventive feedback \citep[eq.~\ref{eqn:zetagrav},
and][]{fau11}.  The cyan line shows a running median; this curve is
reproduced in the other panels for comparison.  The upper and lower right
panels show models where we assume $\eta=2$ and $v_{\rm wind}=680$~km/s
and 340~km/s (cw and sw), respectively.  The lower left panel shows
momentum-driven wind scalings~\citep[vzw;][]{mur05}, in which $\eta\propto
M_{\rm halo}^{-1/3}$ and the outflow speed is $\propto M_{\rm halo}^{1/3}$
(roughly).  Deviations from the no-wind case represent the influence of
other terms on $\dot{M}_{\rm in}$, particularly $\dot{M}_{\rm recyc}$
at high masses and $\zeta_{\rm winds}$ at low masses.
}
\label{fig:ssfrhalo}
\end{figure}

\begin{itemize}

\item $\zeta_{\rm photo}$ represents suppression of inflow owing
to photo-ionisation heating.  This operates at low masses, and
approaches zero below a {\it photo-suppression mass} that increases
from a halo mass of $M_\gamma\sim 10^{8}M_\odot$ during reionisation
to $M_\gamma\sim {\rm few}\times 10^{9}M_\odot$ at the present
epoch~\citep{gne00,oka08}.

\item $\zeta_{\rm quench}$ is associated with whatever physical
process(es) quench star formation in massive halos and prevents
cooling flows, probably related to feedback from supermassive black
holes~\citep[e.g.][]{som08}.  It drops to zero above the {\it quenching
mass} $M_{\rm q}\sim 10^{12}M_\odot$~\citep[e.g.][]{cro06,gab11}, which
may be higher at high-$z$~\citep{dek09}.  This may further depend on the
merger history of galaxies~\citep{hop08}.

\item $\zeta_{\rm grav}$ reflects suppression of inflow by ambient
gas heating owing to gravitational structure formation via the
formation of virial shocks.  Using hydrodynamic simulations with no
outflows, \citet{fau11} determined
\begin{equation}\label{eqn:zetagrav}
\zeta_{\rm grav} \approx 0.47\Bigl(\frac{1+z}{4}\Bigr)^{0.38} 
\Bigl(\frac{M_{\rm halo}}{10^{12}M_\odot}\Bigr)^{-0.25}.
\end{equation}
We note that those simulations did not include metal-line cooling, which
may be an important effect for heating gas in virial shocks~\citep{dek06,ocv08};
nonetheless, we show in Figure~\ref{fig:ssfrhalo} that our simulations
including metal-line cooling yield similar results.

\item $\zeta_{\rm winds}$ is associated with additional heating of
surrounding gas provided by energetic input from winds.  This tends to
affect lower-mass systems more, but is highly dependent on the physics of
how outflows interact with surrounding gas, which is poorly understood.
Recent results~\citep{opp10,vdv11,fau11} have demonstrated that this
can be a significant effect in plausible (though somewhat extreme)
wind models.  Note that although both $\zeta_{\rm winds}$ and $\eta$
arise from winds, the former is a preventive feedback parameter, whereas
$\eta$ is an ejective feedback parameter; the two are not necessarily
related in a simple way, so we keep them separate.

\end{itemize}

Multiplying these terms together, we obtain $\zeta(M_{\rm halo})$ that
is schematically illustrated in Figure~\ref{fig:zeta}.  It is small
at low and high halo masses owing to photo-suppression and quenching
respectively, and approaches unity at intermediate halo masses which is
where vigorous star formation can occur.  This depiction is intended
only to illustrate broad trends, as the actual values and functional
forms of the various $\zeta$ terms are at best only qualitatively known.
The general shape of this curve with a cutoff in accretion at low and
high masses is long known~\citep[e.g.][]{tho96,ker05,cro06}, although the
quantitative masses for the cutoffs are debated to this day~\citep[for
recent work on this see][]{bou10,cat11}.

There may be other sources of energetic preventive feedback that
retard accretion such as cosmic rays, stellar winds, quasar outflows,
local photo-ionisation, and magnetic fields, but their importance
for global galaxy evolution has not yet been firmly established.
There may also be subtle ``amplification" effects by which two or
more preventive mechanisms serve to strengthen each other beyond their
individual impact~\citep{paw09,fin11b}.  Hence the list of individual
$\zeta$'s above is intended to illustrate the sort of physical processes
contributing to preventive feedback, and how they might manifest in the
overall shape of $\zeta$.  The actual trend of $\zeta(M_{\rm halo})$
may involve more complicated and subtle effects than described here.

Figure~\ref{fig:ssfrhalo} illustrates the impact of these various
inflow and feedback terms on the specific star formation rate
(sSFR$\equiv$SFR$/M_*$) as a function of $M_{\rm halo}$, using
large-scale cosmological hydrodynamic simulations with various outflow
models~\citep[see][for details]{dav11}.  The upper left panel shows
the case without outflows.  If all gas entering into the halo ended
up in the ISM, i.e. $\dot{M}_{\rm in}=\dot{M}_{\rm grav}$, then the
relation would be as shown by the solid line, having a positive slope.
Even without any feedback, gravitational heating results in a negative
slope (dashed line).  The results from our simulations are in good
agreement with Equation~\ref{eqn:zetagrav} \citep[from][]{fau11}, since
the simulations themselves are quite similar; the main difference is that
ours include metal-line cooling, but this is not very important until hot
gaseous halos form at $M_{\rm halo}\ga 10^{12}M_\odot$~\citep{ker05,gab11}
since cold accretion is generally limited by the infall time rather than
the cooling time.

The simulations in the two right panels assume $\eta=2$.  If
$\dot{M}_{\rm in}$ is unchanged, then one expects both the SFR and
$M_*$ to be lowered by a factor of three (eq.~\ref{eqn:sfr}),
resulting in no change in sSFR from the no-wind case (the no-wind
cyan curve is reproduced in all panels for comparison).  But clearly
there is a change, which reflects the impact of outflows on
$\dot{M}_{\rm in}$.  At large masses, wind recycling ($\dot{M}_{\rm
recyc}$) returns material to the galaxy rapidly once the wind speed
(680~km/s in the upper right, 340~km/s in the lower right) drops
below the escape velocity~\citep{opp10}.  Hence at large masses
$\eta$ is effectively 0~\citep{fin08}, and sSFR jumps owing to
$\dot{M}_{\rm recyc}$.  For the case of the slower 340~km/s winds
(lower right), the jump occurs at a factor of eight lower in mass
as expected from the factor of two difference in wind speeds.  Winds
also affect sSFR at low masses, where it is suppressed relative to
no winds.  This reflects $\zeta_{\rm winds}$, which is as expected
stronger in the case of the higher wind speed.  In the lower left
we show simulations using momentum-driven wind scalings of
(approximately) $\eta\propto M_{\rm halo}^{-1/3}$.  This flattens
sSFR$(M_{\rm halo})$, in addition to exhibiting different behaviors
for wind recycling and suppression.  

These examples illustrate how star formation rates at a given mass
in simulations are impacted by the various ejective and preventive
feedback processes described above.  For instance, the fact that
observations find no sudden increase in sSFR at any characteristic
mass~\citep[e.g.][]{sal07,dad07} suggests that galaxies do not eject
material at a characteristic wind velocity.

\citet{bou10}, \citet{dut10}, and \citet{kru11} also present empirical
models based on accretion-driven star formation, focusing on the
form and evolution of the sSFR.  Interestingly, \citet{bou10}
demonstrates that a model in which $\zeta=1$ for $10^{11}<M_{\rm
halo}<10^{12.2} M_\odot$ and zero elsewhere nicely reproduces some
key observed properties of high-redshift star-forming galaxies,
including the observed lack of sSFR amplitude evolution from $z\sim
7\rightarrow 2$.  Simulations, in contrast, predict that sSFR evolution
tracks the accretion rate, and thus continues to rise to high
redshifts~\citep{dav08}; analytic models that do not include such
a low-mass cutoff show similar behavior~\citep{dut10}.  If this
low-mass cutoff for accretion were true, it would suggest that there
are additional feedback processes affecting small systems beyond
what is shown in Figure~\ref{fig:zeta}, namely, metagalactic
photo-ionisation.  An alternative explanation that does not employ
a sharp mass cutoff but still reproduces the observed sSFR behavior
is to invoke an observationally-motivated metallicity-dependent
star formation law, as explored by \citet{kru11}.  Pushing such
observations of sSFR out to higher redshifts and lower masses~\citep[e.g.
with the Cosmic Assembly Near-Infrared Deep Legacy Survey;][]{gro11,koe11}
should provide interesting constraints on the physical mechanisms
regulating early, low-mass galaxy growth.

\section{Gas fractions}\label{sec:mgr}

%\begin{figure}
%\vskip -0.5in
%\setlength{\epsfxsize}{0.65\textwidth}
%\centerline{\epsfbox{fig.fgequil.ps}}
%\vskip -3.1in
%\caption{Gas fraction evolution predicted by Equation~\ref{eqn:fgas},
%assuming $\tdep\propto t_H$ and sSFR evolution is governed solely by
%$\dot{M}_{\rm grav}$ (eq.~\ref{eqn:Min}).  We show results using three forms
%for $\dot{M}_{\rm grav}$ for comparison, from \citet[solid black]{dek09},
%\citet[dashed green]{fak10}, and \citet[dotted blue]{bou10}; the
%differences are not large.  Curves are arbitrarily normalized at
%$z=0$ to sSFR$=0.1$~Gyr$^{-1}$ and $\tdep=2$~Gyr; the redshift
%dependence is insensitive to this.  Molecular (CO) gas
%fraction measurements are shown from the samples of \citet[red points at
%$z=1.2,2.3$]{tac10} and \citet[blue point at $z=0.4$]{gea11}.
%}
%\label{fig:fgequil}
%\end{figure}

A galaxy's gas fraction is defined here as
\begin{equation}\label{eqn:fgas}
\fgas \equiv \frac{\mgas}{\mgas+M_*} = \frac{1}{1+(\tdep {\rm sSFR})^{-1}}
\end{equation}
where in the second equality we have employed the {\it depletion time}
$\tdep\equiv \mgas/$SFR.  We argue below that the latter formulation offers the
intuitive advantage that it splits $\fgas$ into a term that is fairly
insentive to feedback ($\tdep$) and a term that depends strongly on
feedback (sSFR).

The depletion time measures the timescale over which gas, when
present in the ISM, gets converted into stars.  This is expected
to be primarily determined by the star formation law, such as the
observed \citet{ken98} relation between gas surface density
($\Sigma_{\rm gas}$) to SFR surface density ($\Sigma_{\rm SFR}$).
Indeed, simulations by \citet[][hereafter DFO11; see their
Figure~4]{dav11b} assuming a Kennicutt-Schmidt law show that $\tdep$
is essentially independent of outflows, and scales as
\begin{equation}
\tdep\propto t_H M_*^{-0.3},
\end{equation}
where $t_H$ is the Hubble time.

We can derive this scaling of $\tdep$ directly from the star
formation law.  The temporal scaling can be most easily understood
using a formulation of the star formation law given by SFR$\approx
0.02 \mgas/t_{\rm dyn}$, where $t_{\rm dyn}$ is the dynamical time of
the star formation region~\citep[e.g.][]{sil97,kru07,gen10}.  This then
gives $\tdep\propto t_{\rm dyn}$, which in a canonical disk model scales
as the Hubble time $t_H$~\citep{mo98}.  Meanwhile, the stellar mass
dependence arises from the Kennicutt law plus the ISM gas profile.
The Kennicutt law states that $\Sigma_{\rm SFR}\propto \Sigma_{\rm
gas}^N$ (where $N\approx 1.4$), from which it is straightforward to
show that $t_{\rm dep}\propto \Sigma_{\rm gas}^{1-N}$.  In simulations,
$\Sigma_{\rm gas}\propto M_*^{3/4}(1+z)^{2}$~(DFO11), which gives rise
to the weak anti-correlation with $M_*$ quoted above.  We caution that
these dependences may be somewhat different in the real Universe since
these simulations lack the resolution to properly model the internal
structure of galaxies.

The evolution of $\fgas$ depends on the evolution of $\tdep$ and
sSFR.  The former evolves with $t_H$ (e.g. as $(1+z)^{-1.5}$ in the
matter-dominated regime), while the latter is generally driven by
cosmic inflow (eq.~\ref{eqn:Min}) which scales as $(1+z)^{2.25}$ if
driven by gravitional infall.  Combining these, galaxy gas fractions
are predicted to evolve with time roughly as $\sim (1+z)^{2.25}t_H$
(if $\fgas$ is not near unity), which increases slowly with redshift,
qualitatively consistent with observations~\citep{tac10,gea11}.  Hence in
the equilibrium scenario, galaxy gas fractions represent a competition
between supply and consumption, such that {\it galaxies become less
gas-rich with time because the gas supply rate drops faster than the
gas consumption rate.}

\section{Metallicities}\label{sec:mzr}

The global metallicity within the ISM is given by the enrichment
rate, which is the yield $y$ times SFR, divided by the mass inflow
rate $\dot{M}_{\rm in}$ that must be enriched.  As derived in
\citet{fin08}, if the inflow is pre-enriched there is an additional
term that depends on $\alpha_Z\equiv Z_{\rm in}/Z_{\rm ISM}$, where
$Z_{\rm in}$ and $Z_{\rm ISM}$ are the metallicities of the inflowing
and ambient ISM gas, respectively:
\begin{equation}\label{eqn:mzr}
Z_{\rm ISM} = y\frac{{\rm SFR}}{\dot{M}_{\rm in}} = \frac{y}{1+\eta} \frac{1}{1-\alpha_Z}.
\end{equation}
Hence the mass-metallicity relation and its evolution are established
by a the mass and redshift dependence of $\eta$ and $\alpha_Z$.  In
our currently favored outflow model~\citep{dav11}, $\eta$ has a
significant mass dependence but little or no redshift dependence,
while $\alpha_Z$ is generally small but has a significant redshift
dependence.  Equation~\ref{eqn:mzr} would then suggest that the
shape of the mass-metallicity relation is primarily established by
$\eta(M_*)$, while its evolution is driven by $\alpha_Z$; this was
demonstrated for simulations in DFO11.  Hence in this
scenario, {\it the shape of the mass-metallicity relation is modulated
by the fraction of inflow that forms stars, while its evolution is
governed by the enrichment level of infalling gas.}

Conspicuously absent in this scenario is any explicit reference to
potential wells of galaxies, or any consideration of outflow
velocities versus escape velocities.  These processes are canonically
believed to govern the mass-metallicity
relation~\citep[e.g.][]{dek86,tre04}; the phrase ``metals can more
easily escape from the shallower potential wells of small galaxies"
is oft-repeated.  However, in our scenario, it is instead the net
mass outflow rate that is the key determinant of the mass-metallicity
relation, and the potential well depth is at most only indirectly
implicated.

Since the vast majority of metals in the IGM are deposited there
by outflows~\citep[e.g.][]{opp08,opp11}, the infalling gas metallicity
is a direct measure of $\dot{M}_{\rm recyc}$.  $Z_{\rm in}$ is given
by the metal mass arriving in the form of recycled winds, divided
by the total mass inflow rate, i.e.
\begin{equation}\label{eqn:Zin}
Z_{\rm in} = Z_{\rm recyc} \frac{\dot{M}_{\rm recyc}}{\dot{M}_{\rm recyc}+\zeta \dot{M}_{\rm grav}},
\end{equation}
where the denominator is $\dot{M}_{\rm in}$ from Equation~\ref{eqn:Minsplit}.

Under the typical case of highly mass-loaded outflows, the outflowing
metallicity must be similar to the ambient ISM metallicity, i.e.
$Z_{\rm out}\approx Z_{\rm ISM}$.  Furthermore, since galaxies
evolve slowly in metallicity~\citep[e.g.][DFO11]{bro07} and wind
recycling times are typically of order a Gyr~\citep{opp10}, the
galaxy metallicity has probably not evolved strongly from when the
gas was ejected to when it is being re-accreted, and hence $Z_{\rm
recyc}\approx Z_{\rm out}$.  Substituting $Z_{\rm recyc}=Z_{\rm
ISM}$ into Equation~\ref{eqn:Zin} and solving for $\dot{M}_{\rm
recyc}$ yields
\begin{equation}\label{eqn:Mrecyc}
\dot{M}_{\rm recyc} = \frac{\alpha_Z}{1-\alpha_Z} \zeta \dot{M}_{\rm grav}.
\end{equation}
This relates the mass recycling term in the inflow equation
(eq.~\ref{eqn:sfr}) to the metallicity infalling into the ISM.
The advantage of formulating recycling in terms of 
$\alpha_Z$ is that it is in principle an observable quantity via absorption
or emission measures in the outskirts of galaxies.  In contrast,
$\dot{M}_{\rm recyc}$ is not directly measurable since it is not
clear how to distinguish recycled wind inflow from other inflow,
or even how to measure galaxy inflow rates at all.  Note that since
galaxy metallicities evolve slowly upwards with time,
Equation~\ref{eqn:Mrecyc} will tend to slightly underestimate
$\dot{M}_{\rm recyc}$ for a given $\alpha_Z$.  The ejection of winds
from one galaxy (typically a satellite) being accreted onto another
(typically the associated central) could also affect $\alpha_Z$,
which would also cause an underestimate in $\dot{M}_{\rm recyc}$
since the satellites are generally smaller and hence lower metallicity.

\section{The Equilibrium Relations \& Implications}\label{sec:equil}

We can substitute Equation~\ref{eqn:Mrecyc} into
Equation~\ref{eqn:Minsplit} to obtain
\begin{equation}\label{eqn:sfrfull}
{\rm SFR} = \frac{\zeta \dot{M}_{\rm grav}}{(1+\eta)(1-\alpha_Z)}.
\end{equation}
This is the key equation that delineates how galaxy star
formation rates are governed by accretion and feedback processes,
i.e. baryon cycling.  This equation, together with
Equations~\ref{eqn:fgas} and \ref{eqn:mzr}, 
represent the {\it equilibrium relations} that govern the stellar,
gas, and metal content of galaxies across cosmic time.  Galaxies
will tend to lie around these relations owing to a balance of inflow,
outflow, and star formation.

The equilibrium relations depend on three parameters: $\eta$,
$\zeta$, and $\alpha_Z$, representing ejective feedback (i.e.
outflows), preventive feedback, and wind recycling.  Additionally,
the star formation law governs $\tdep$, $\dot{M}_{\rm grav}$ is set
by cosmology, and $y$ is set by nucleosynthetic processes.  Assuming
those are well-established, the mass and redshift (and possibly
environmental) dependence of $\eta$, $\zeta$, and $\alpha_Z$ govern
the evolution of the global SFR, $\fgas$, and $Z_{\rm ISM}$ of
galaxies.  Note that since the mass and redshift dependence of
these parameters are not fully known, the actual number of free
parameters can be significantly larger than three.

There are many possible ways to characterize simulation results into an
analytic formalism~\citep[e.g.][]{nei11}.  One virtue of our particular
parameterization is that the parameters involved are, at least in
principle, directly observable.  This provides an optimally direct
connection from observations to constraints on galaxy formation models.
Unfortunately, measuring these parameters is challenging, but preliminary
constraints have already been obtained.

For instance, $\eta$ has been constrained in high-$z$ galaxies to have a
value of order unity or more~\citep[e.g.][]{ste10,gen11}.  $\alpha_Z$
can be constrained by examining metallicities in the outskirts of
low-$z$ galaxies~\citep[e.g.][]{bre09,mor11}.  Constraining $\zeta$
by direct observations would require an accurate census of all halo
gas which is highly challenging, but aside from $\zeta_{\rm winds}$,
its main terms can be constrained using a combination of relatively
straightforward numerical work ($\zeta_{\rm photo}$ and $\zeta_{\rm
grav}$) and empirical arguments ($\zeta_{\rm quench}$).

The equilibrium relations have some interesting implications for
the behavior of SFR, $\fgas$, and $Z$.  For instance, hydrodynamic
simulations indicate that the star formation history of galaxies is
insensitive to the assumed star formation law~\citep{kat96,sch10}.
This seems paradoxical at first, but is straightforwardly seen from
Equation~\ref{eqn:sfrfull}, since there is no dependence here (or in
the metallicity equation) on the star formation law.  The star formation
law only affects the gas fractions, via $\tdep$.  This can be regarded
as a self-regulation mechanism~\citep{sch10}, in which gas collects in
galaxies as required in order to achieve the star formation rate set by
the balance of inflows and outflows.

Another straightfoward prediction of the equilibrium relations
is that if one desires the mass-metallicity relation to scale as
$Z\propto M_*^{1/3}$ at small masses as observed~\citep{tre04,lee06},
then Equation~\ref{eqn:mzr} directly implies $\eta\propto M_*^{-1/3}$
(assuming $\alpha_Z\ll 1$), roughly as expected for momentum-driven
winds~\citep{mur05}.  Indeed, simulations assuming such a scaling
appear to provide a good match to mass-metallicity relation
observations~\citep[][DFO11]{fin08}.

It is instructive to combine Equations~\ref{eqn:mzr} and \ref{eqn:sfrfull}
to give
\begin{equation}
\zeta = \frac{\rm SFR}{\dot{M}_{\rm grav}} \frac{y}{Z_{\rm ISM}}
\end{equation}
The first ratio is the {\it halo star formation efficiency} (SFE), i.e.,
the fraction of gravitational infall into a halo that ends up forming
into stars\footnote{This is distinguished from the ISM SFE, which is how
much gas is converted into stars over some characteristic galaxy
timescale, or the ``cosmological" SFE, which is the galaxy stellar mass
divided by the cosmologically-expected halo baryon mass 
(i.e. $M_*/f_bM_{\rm halo}$).}, 
while the second ratio quantifies the metal retention
fraction within galaxies.  If the halo mass and metal yield can be
determined, measuring SFR$/Z_{\rm ISM}$ provides a 
quantitative constraint on $\zeta$.  This can be done at least at $z\sim 0$ 
with existing data from e.g. the Sloan Digital Sky Survey.

\section{A Sample Equilibrium Model}\label{sec:sample}

The equilibrium model can be used to quickly explore parameter space
and obtain intuition about the governing physics for galaxy properties
of interest.  We illustrate this here by presenting results for the
evolution of galaxies in a full equilibrium model.

\begin{figure}
\vskip -0.4in
\setlength{\epsfxsize}{0.6\textwidth}
\centerline{\epsfbox{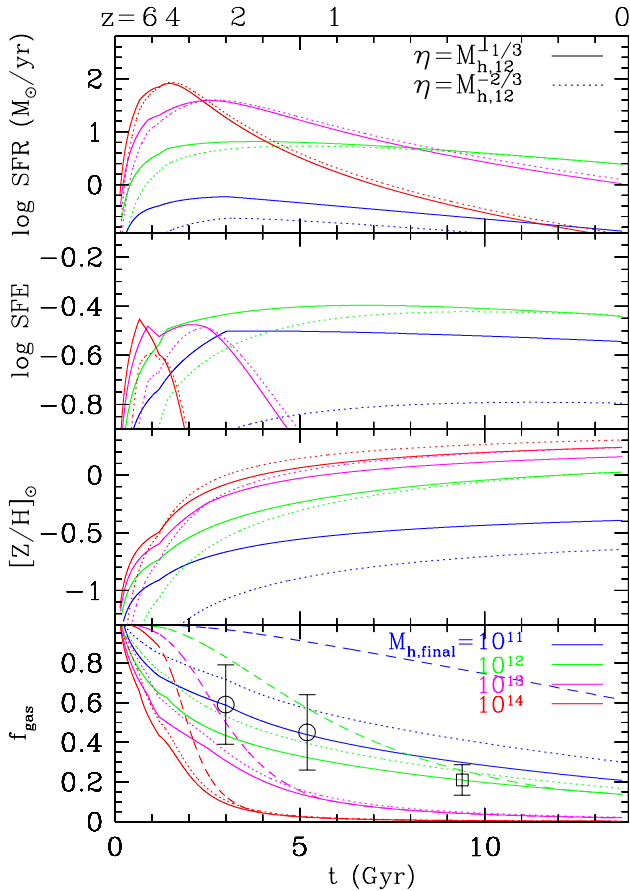}}
\vskip -0.5in
\caption{Equilibrium model evolution of four galaxies that have
final $z=0$ halo masses of $10^{11} M_\odot$, $10^{12} M_\odot$,
$10^{13} M_\odot$, and $10^{14} M_\odot$ (blue, green, magenta, and
red lines, respectively).  Panels from top to bottom show the star
formation rate, cosmic star formation efficiency defined as the
fraction of baryons entering the halo that form into stars, gas
fractions, and metallicities.  Solid lines show $\eta=(M_{h}/10^{12}
M_\odot)^{-1/3}$, and dotted lines show $\eta=(M_{h}/10^{12}
M_\odot)^{-2/3}$.  The star formation histories peak earlier and
at higher SFRs for larger galaxies, reminiscent of the empirical
``staged galaxy formation" model of \citet{noe07}.  Star formation
efficiencies are typically $\sim 1/2-1/3$ for star-forming galaxies,
and drop quickly about the quenching halo mass of $10^{12.3}M_\odot$
for more massive systems.  Galaxies self-enrich to above one-tenth
solar very early on, and then evolve slowly in metallicity.  Observed
gas fractions for fairly massive galaxies from \citet{tac10} at
$z=1.2$ and 2.2 and \citet{gea11} at $z=0.4$ are indicated, to be
compared with predictions for $\sim 10^{12-13}M_\odot$ halos; the
model predictions are generally too low.  The dashed lines show the
experiment of adding a metallicity dependence of $\propto Z^{-2}$
to $\tdep$ when sub-solar, which keeps early galaxies gas-rich and
improves agreement with data.
}
\label{fig:equil}
\end{figure}

Figure~\ref{fig:equil} shows the evolution of the SFR, halo SFE,
$Z$, and $\fgas$ for four galaxies spanning the indicated range of
final ($z=0$) halo masses.  These are computed using
Equations~\ref{eqn:fgas}, \ref{eqn:mzr}, and \ref{eqn:sfrfull},
tracking the stellar and halo mass growth starting at an early epoch
when the halo is at the photo-suppression mass.  We take $\dot{M}_{\rm
grav}$ from Equation~\ref{eqn:Min}, parameterize $\tdep=0.4 t_H
(M_*/10^{10} M_\odot)^{-0.3}$ as discussed in \S\ref{sec:mgr}, and
take $\alpha_Z=(0.5-0.1z)(M_*/10^{10}M_\odot)^{0.25}$ (with
$\alpha_Z\geq0$) as a crude parameterization of simulation results
from DFO11.  We choose $\zeta$ as described in
Figure~\ref{fig:zeta}, and define $\eta$ as indicated in the upper
right of the Figure: solid lines approximately represent momentum-driven
wind scalings, while dotted lines represent energy-conserving wind
scalings.  We also include instantaneous recycling of 18\% of star
formation back into gas as expected for a \citet{cha03} IMF, but
do not include further stellar mass loss.  We reiterate that these
``base model" parameter choices are at some level arbitrary, and
are intended only to illustrate how parameter variations influence
observables.

The green line represents a Milky Way-sized halo of $10^{12} M_\odot$.
At $z=0$, it has SFR$\approx 2.5 M_\odot$/yr, $M_*\approx 5\times
10^{10} M_\odot$, $\fgas\approx 0.1$, and $Z\approx Z_\odot$, in
fair agreement with measured values and showing that our parameter
choices are reasonable.  Larger galaxies form stars more vigorously
at earlier epochs and for shorter intervals, which is qualitatively
similar to the behavior in the empirical ``staged" galaxy formation
model of \citet{noe07}.  The peak SFRs are 50--100~$M_\odot/$yr at
$z\sim 2-3$, which is lower than the observed values for the largest
main sequence galaxies at that epoch by a factor of a few, reiterating
the issue noted in simulations by \citet{dav08} that observed galaxy
SFRs at that epoch approach or exceed their cosmic accretion rate;
the resolution to this quandary remains unclear~\citep[see e.g.][for
ideas]{bou10,kru11}.

The halo SFE is plotted in the second panel.  For star-forming galaxies,
this efficiency is roughly one-third to one-half over most of cosmic time.
Higher mass halos show a marked drop in efficiency once they grow above
the quenching mass, here assumed to be $2\times 10^{12} M_\odot$ at
all epochs.  Going to a steeper scaling of $\eta(M_h)\propto M_h^{-2/3}$
(dotted lines) more strongly suppresses low-mass galaxy growth, which
is favored in semi-analytic models to reproduce the observed faint-end
slope of the stellar mass function~\citep[e.g.][]{som08}.

Metallicity evolution is very rapid in all galaxies early on, and
galaxies typically enrich to $\ga 0.1Z_\odot$ within the first Gyr,
as seen in simulations~\citep{dav06,fin11}.  Evolution is slow
thereafter.  Galaxies tend to follow parallel tracks, showing that
the shape of the mass-metallicity relation does not evolve
much~(DFO11).  A steeper scaling of $\eta(M_h)$ strongly
steepens the mass-metallicity relation in smaller galaxies at all
redshifts, as expected from Equation~\ref{eqn:mzr}.

$\fgas$ drops slowly with time since the accretion rate drops faster
than the consumption rate, as discussed in \S\ref{sec:mgr}.  Here
we include some observational comparisons (large symbols), from CO
measurements by \citet{tac10} and \citet{gea11}.  As \citet{tac10}
noted, the relatively slow evolution requires continual replenishment,
as our model naturally predicts.  However, the $z\sim 1-2$ gas
fractions are higher than the model predictions, because the observed
galaxies are best compared to the green and magenta lines representing
final halo masses of $\sim 10^{12}-10^{13} M_\odot$.  The equilibrium
model can quickly test whether parameter variations can reconcile 
this discrepancy.

According to Equation~\ref{eqn:fgas}, higher $\fgas$ values can be
achieved by raising sSFR and/or $\tdep$.  Raising sSFR e.g. by lowering
$\eta$ would concurrently alleviate the discrepancy with observed $z\sim
2$ SFRs, but would likely overproduce stars globally.  Delaying star
formation by accumulating gas could explain the discrepancy, though
it seems somewhat contrived to have galaxies at all masses suddenly
start vigorously consuming gas around $z\sim 2$.  A steeper scaling
of $\eta(M_h)$ works in this direction, but the dotted lines show
that this only helps marginally; a redshift-dependent $\eta$ that
is higher at high-$z$ may work better, but the physical motivation
is not obvious.  Another avenue is to raise $\tdep$ by appealing
to the observed lower gas consumption rate in low-metallicity
systems~\citep{bol11}.  An illustration of this is shown as the dashed
lines in Figure~\ref{fig:equil}, where we add a dependence of $Z^{-2}$
to $\tdep$ when below solar metallicity.  Within this model, we find that
this steep metallicity dependence is required in order to sufficiently
raise gas fractions at high-$z$; however, it is not physically-motivated,
and does not necessarily reflect recently proposed metallicity-dependent
star formation laws~\citep[e.g.][]{kru11}.  The delay in star formation
results in higher gas fractions at $z\sim 2$ in better agreement with
data, but at lower redshifts when metallicities approach solar this makes
less difference.  Another avenue explored by \citet{bou10} is to postulate
that star formation can only occur in halos above $10^{11}M_\odot$,
which has the advantage of producing higher has fractions in smaller
systems at earlier epochs; they demonstrate that this can broadly
match the observed evolution of galaxy gas fractions.

This exercise illustrates how an equilibrium model can be utilized
to gain intuition about various physical processes in galaxy
evolution.  The relatively small number of parameters and their
direct connection to observable physical processes makes this type
of model easier to interpret than modern simulations or semi-analytic
models.

\section{Departures From Equilibrium}\label{sec:scatter}

Stochastic variations in $\dot{M}_{\rm in}$, including mergers, can
cause departures from equilibrium.  This generates scatter about
the equilibrium relations.  Generically, departures from equilibrium
tend to return galaxies towards equilibrium. This self-regulating
behavior is why we dub this scenario the equilibrium model.

To illustrate this behavior, consider a galaxy experiencing an
upward fluctation in $\dot{M}_{\rm in}$.  Its $\fgas$ increases,
and perhaps $M_*$ as well if there is a small galaxy accompanying
the infall, while $Z_{\rm ISM}$ decreases since either fresh infall
or a lower-mass galaxy will have lower metallicity.  This moves the
galaxy off the equilibrium relations.  The increased gas content
immediately stimulates more vigorous star formation, which over
time enriches the galaxy as it consumes the excess gas.  This then
lowers its gas content and increases its metallicity, moving the
galaxy back towards equilibrium.

Conversely, a temporary lull in accretion will make a galaxy more
gas-poor and metal-rich as it consumes its existing gas.  Quantitatively,
its metallicity will evolve following the relation $\Delta Z_{\rm
ISM}\approx y\Delta M_*/M_{\rm gas}$~\citep[see eq.~19 of][in the
case of zero accretion]{fin08}, which is steeper than the MZR and
hence will move the galaxy above the MZR.  In time, hierarchical
growth will bring in fresh gas, lowering the metallicity and
increasing the gas content to return the galaxy towards equilibrium.
In this way, galaxies oscillate around the equilibrium relations,
constantly being perturbed from them and driven back owing to
fluctuations in infall.

An inevitable prediction of this scenario is that departures from
the equilibrium relations will correlate with star formation rate,
gas fraction, and metallicity.  From the above scenarios, one can
see that {\it at a given mass, galaxies that are gas-rich (gas-poor)
and metal-poor (metal-rich) will have higher (lower) star formation
rates.}  These trends are qualitatively consistent with
observations~\citep{ell08,lar10,man10,pee10}.

This trend is illustrated for gas fractions and star formation rates
by the coloured points in Figure~\ref{fig:ssfrhalo}, showing that
at a given mass, high SFR and high $\fgas$ go hand in hand.  Figure~1
of DFO11 analogously shows that high SFR accompanies low $Z_{\rm
gas}$ at a given $M_*$.  Note that these second-parameter trends
do not arise from outflows, being present even in simulations without
winds.  Instead, it is a direct and unavoidable consequence of
equilibrium, and results from galaxies' self-regulating response
to fluctuations in inflow rather than any feedback process.

Quantitatively, the scatter around the equilibrium relations depends
on how quickly galaxies can return to equilibrium after being
perturbed.  To return, there must be sufficient infall to re-equilibrate
the galaxy.  The timescale for this to happen can be quantified by
the dilution time~\citep{fin08}:
\begin{equation}\label{eqn:tdil}
t_{\rm dil}\equiv\frac{M_{\rm gas}}{\dot{M}_{\rm in}}=(1+\eta)^{-1} \frac{M_{\rm gas}}{\dot{M}_*}=(1+\eta)^{-1} t_{\rm dep}.
\end{equation}
If the dilution time is small compared to the inflow fluctuation
timescale, then the scatter will be small~\citep{fin08}.  The 
inflow fluctuation timescale likely depends on mass and environment,
with small galaxies generally suffering (relatively) larger perturbations.
Equation~\ref{eqn:tdil} shows that the dilution time depends
on $\eta$ and $\tdep$, and hence observations of the scatter versus
$M_*$ provides an independent constraint on $\eta(M_*)$ and
$\tdep(M_*)$.

\citet{dut10} argues, based on an analytic model of accretion-driven
galaxy formation similar to this one, that the scatter in the
observed $M_*-$SFR relation must be driven by fluctuations in inflow,
since assuming zero scatter in the relation between $\dot{M}_{\rm
in}$ and $M_{\rm halo}$ results in a relation that is too tight
compared to observations.  However, hydrodynamic simulations that
implictly include inflow fluctations also yield a small
scatter~\citep[e.g.][]{dav08,fin11}.  Therefore it is not solely
inflow fluctuations that govern the scatter, it is the more complex
relationship between inflow fluctuations and the dilution time.
Equation~\ref{eqn:tdil} suggests that this, in turn, depends on the
outflow rate ($\eta$) and gas consumption rate ($\tdep$).  The fact
that feedback regularizes galaxy properties may have other effects
on galaxy properties.  For instance, it has long been suggested
that the low scatter in the Tully-Fisher relation arises owing to
feedback processes, since fluctuations in halo growth alone would
naively predict a scatter that is large compared to data~\citep{eis96}.

Satellite galaxies lie permanently off the equilibrium relations,
because the inflowing filaments bypass them and flow to the centers
of halos.  Hence they are expected to end up with lower gas content
and higher metallicities than centrals of the same mass; this is as
observed~\citep{pee09}.  How far they lie off the equilibrium relations
depends on their gas reservoir at the time they are cut off from their
supply, which simulations indicate is typically $\sim 1$~Gyr after falling
into a hot gas-dominated halo~\citep{sim09}.  Note that some satellite
galaxies fall in along the inflowing filaments, and in those cases they
are not bypassed since they are actually part of the inflow.  However,
these satellites are expected to quickly merge into the central galaxy,
and so will not typically end up as part of the long-lived satellite
population.

Major mergers are another population lying far out of equilibrium.
It is not merely that such systems represent a particularly large
inflow perturbation, it is that the induced torques drive gas flows
that fuel central star formation~\citep{mih96} making cosmological
inflow mostly irrelevant during the merger event.  In a global
context, major mergers are seen to be responsible for only a small
fraction of overall cosmic star formation~\citep[e.g.][]{jog09},
although they may be where much of the central black hole mass
growth occurs~\citep[e.g.][]{dim05}.  Overall, the equilibrium model
is broadly valid for central galaxies that are quiescently forming
stars, i.e. main-sequence galaxies~\citep{noe07}, that dominate
cosmic star formation.

\section{Before Equilibrium: The Gas Accumulation Phase}\label{sec:zeq}

Since the infall rate has a steeper redshift dependence than the
consumption rate (e.g. $\sim (1+z)^{2.25}$ vs. $t_H^{-1}$), at
sufficiently early epochs galaxies will be in a {\it gas accumulation
phase} during which galaxies cannot process gas into stars as fast as
they receive it~\citep{bou10,kru11}.  Only when consumption can keep up
with supply will equilibrium be achieved.  During the gas accumulation
phase, gas fractions are expected to be higher and metallicities lower
than predicted by equilibrium.

\begin{figure}
\vskip -0.4in
\setlength{\epsfxsize}{0.65\textwidth}
\centerline{\epsfbox{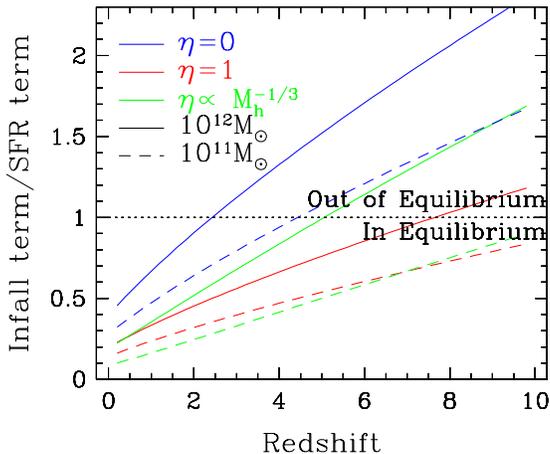}}
\vskip -3.2in
\caption{Evolution of the ratio of the infall term (eq.~\ref{eqn:sfr})
versus the SFR term (SFR=$0.02 M_{\rm gas}/t_{\rm dyn}$), for
different choices for $\eta$ and halo mass (solid and dashed lines
represent $M_h=10^{12}$ and $10^{11}M_\odot$, respectively).  This
calculation assumes $\fgas=0.5$ for illustration purposes, which
is typical at high-$z$ but in detail varies with $\eta$ and $M_h$.
With no outflows ($\eta=0$; blue lines), equilibrium for massive
galaxies occurs at $z\sim 2$, while less massive galaxies equilibrate
earlier.  With modest outflows ($\eta=1$; red lines), equilibrium
is already pushed back into the early Universe.  The green lines
assume momentum-driven wind scalings for $\eta$ from \citet{opp08},
which pushes equlibrium to $z\sim 5$ for massive galaxies, and much
earlier for smaller ones.
}
\label{fig:zeq}
\end{figure}

We can estimate the redshift $z_{\rm eq}$ where gas accumulation
ends and equilibrium is attained.  Star formation must be able to
occur fast enough to satisfy SFR$= \dot{M}_{\rm in}/(1+\eta)$
(eq.~\ref{eqn:sfr}).  We assume that at early times, $\zeta\approx
1$ and $\alpha_Z\approx 0$, so that $\dot{M}_{\rm in}\approx
\dot{M}_{\rm grav}$ (eq.~\ref{eqn:Min}).  We further assume that
SFR=$0.02 M_{\rm gas}/t_{\rm dyn}$, where $t_{\rm dyn}\approx 10^8
(1+z)^{-3/2}$~yr is the disk dynamical time.  Taking
$M_{\rm gas}\approx \fgas f_b M_{\rm halo}$, we obtain an equation
for $z_{\rm eq}$:
\begin{equation}\label{eqn:zeq}
1+z_{\rm eq}\approx
\Bigl[5 \fgas (1+\eta) \Bigr]^{4/3} \Bigr(\frac{M_{\rm halo}}{10^{12} M_\odot}\Bigl)^{-0.2}.
\end{equation}

We illustrate these trends in Figure~\ref{fig:zeq}.  For sizeable
high-$z$ galaxies with $\fgas\approx 0.4-0.5$~\citep{tac10} and no
outflows ($\eta=0$), we obtain $z_{\rm eq}\approx 1.5-2.5$, in
agreement with~\citet{kru11}.  But the superlinear dependence on
$\fgas$ and $(1+\eta)$ means that the results are quite sensitive
to these values.  For instance, even a modest outflow rate of
$\eta=1$ yields $z_{\rm eq}\approx 5-7$.  Effectively, outflows
lower the amount of inflow that needs to be processed into stars,
allowing for an earlier equilibration epoch.  There is a weak halo
mass dependence as well such that smaller galaxies equilibrate
earlier.  Note that this derivation of $z_{\rm eq}$ depends on the
star formation law: if the star formation law was different in early
galaxies owing e.g. to metallicity effects, then this would also
impact when galaxies achieve equilibrium.  As such, a precise
prediction of $z_{\rm eq}$ is sensitive to poorly known factors.
Nevertheless, with realistic outflows, it is likely that $z_{\rm
eq}\gg 2$, and hence galaxies live in equilibrium over the
vast majority of cosmic time.

Observationally, \citet{pap10} used the star formation rates, masses,
and (estimated) gas contents of high-$z$ Lyman break galaxies to
infer that gas accumulation occurs down to $z_{\rm eq}\sim 4$, after
which accretion and star formation track each other as expected in
equilibrium.  Hence there is some direct empirical support for an
early gas accumulation epoch.  Constraining $z_{\rm eq}$ more
precisely will provide quantitative constraints on gas processing
rates in early galaxies.

\section{Summary and Discussion}\label{sec:summary}

We have presented a simple formalism for understanding the evolution of
the stellar, gaseous, and metal content of galaxies, inspired by
intuition gained from cosmological hydrodynamic simulations.  This
formalism is encapsulated by the equilibrium relations:
\begin{eqnarray} \label{eqn:full}
{\rm SFR} &=& \frac{\zeta \dot{M}_{\rm grav}}{(1+\eta)(1-\alpha_Z)}, \\
\fgas &=& \frac{1}{1+(\tdep {\rm sSFR})^{-1}},\\
Z_{\rm ISM} &=& \frac{y}{1+\eta} \frac{1}{1-\alpha_Z}.
\end{eqnarray}
These relations are established by a balance between inflows and
outflows, and evolve on cosmological timescales over which inflow
and outflow rates slowly vary.  They are primarily governed by three
baryon cycling parameters that describe ejective feedback ($\eta$),
preventive feedback ($\zeta$), and the re-accretion of ejected
material ($\alpha_Z$); each of these parameters is in principle
directly observable, but at present has poorly known dependences
on mass and redshift (and perhaps other properties).
Additionally, they depend on $\dot{M}_{\rm
grav}$, the gravitational infall rate of baryons into the halo set
by $\Lambda$CDM, $\tdep$, the ISM gas depletion time, and $y$, the
metal yield.  These relations capture, to first order, the behavior
of galaxies in modern hydrodynamic simulations that incorporate
these processes dynamically within a hierarchical structure formation
scenario.  

The equilibrium model is broadly valid for quiescently star-forming
central galaxies, at epochs where star formation is able to keep up with
inflow (i.e.  past the gas accumulation epoch; Figure~\ref{fig:zeq}),
and when averaged over timescales longer than stochastic fluctuations
in the inflow rate.  On shorter timescales, galaxies oscillate around
the equilibrium relations such that more (less) rapidly star-forming
galaxies at a given mass having higher (lower) gas fractions and lower
(higher) metallicities.  The scatter about the relations is governed by
a competition between the dilution time $t_{\rm dil}=(1+\eta)^{-1}\tdep$
and the inflow stochasticity timescale.  This model is {\it not} valid
for satellite galaxies disconnected from feeding filaments, or for
galaxies undergoing a major merger where gas feeding is temporarily
driven by internal dynamical processes.  Nonetheless, observations
indicate that quiescently star-forming galaxies along the so-called
galaxy main sequence dominate cosmic star formation at all epochs where
measured~\citep[e.g.][]{noe07,rod11}, and hence this model describes
how the bulk (but not all) of the stars in the Universe formed.

Most current galaxy formation models, both hydrodynamic and semi-analytic,
already include inflow and outflow processes within growing large-scale
structure.  Hence there is no new physics in the equilibrium model.
What is notable is not what this scenario contains, but rather what it
{\it doesn't} contain.  In particular, there is no explicit mention
of mergers, disks, environment, cooling radii, or virial radii---
all central elements in the canonical scenario for galaxy formation.
Such elements are automatically accounted for in hydrodynamic simulations,
which form disks, merge them, and implictly include environmental effects
within growing large-scale structure.  Yet the equilibrium relations well
describe such simulations without reference to these elements, suggesting
that they are not of primary importance for the evolution of galaxies'
SFR, $\fgas$, and $Z$.  

In a broader context, the usefulness of the equilibrium scenario is that
it lays bare the overwhelming complexity of modern galaxy formation
models, and isolates those aspects that are critical for governing
global galaxy evolution, thereby providing a simpler intuitive view
for how galaxies grow.  Although the canonical ``halo-merger" view of
disks cooling within halos and merging to drive galaxy evolution is not
incorrect (i.e.  these processes do happen), such a view obfuscates the
primary driver of global galaxy evolution, namely the balance between
inflows, outflows, and star formation.  Indeed, the very notion of a
galaxy halo, which is central to the classical view of galaxy formation,
is only a second-order effect in the equilibrium scenario: The equilibrium
relations are driven by the total inflow rate, and the ``lumpiness"
of that inflow owing to individual halos merely manifests as scatter
around these relations.

We have argued that the equilibrium model provides a reasonable
description of sophisticated simulations, but this in no way guarantees
that it accurately describes the real Universe.  It is encouraging that
certain unavoidable predictions such as continual gas replenishment and
the second-parameter dependence of the mass-metallicity relation seem
to be in broad agreement with observations.  But much work remains
to be done in order to fully test this scenario.  In particular,
the equilibrium model centrally invokes a continual cycle of baryons
flowing in and out of galaxies as a key moderator of galaxy evolution.
But direct observational evidence for such processes is currently
scant~\citep[see e.g.][]{rub11}.  Critically testing and constraining
these baryon cycling processes, particularly within circum-galactic
gas where such processes are likely to be most prominent, will be a key
contribution from upcoming multi-wavelength observational facilities.

The equilibrium model provides a re-parameterized framework for
understanding certain key governing aspects of galaxy evolution, but is
far from a full solution to the problem.  To fully solve galaxy evolution,
we must at minimum understand the physics that governs $\eta$, $\zeta$,
and $\alpha_Z$, which will require concerted efforts on both observational
and theoretical fronts.  Furthermore, halo accretion rates, metal yields,
and the star formation law remain uncertain, particularly in regimes
such as small low-metallicity galaxies.  Perhaps most importantly, this
scenario in its current form explicitly does not address many interesting
aspects of galaxy evolution such as the establishment of the Hubble
sequence and the growth of central black holes.  It also does not include
processes that may be central to the evolution of certain classes of
galaxies; for instance, it does not account for stellar (``dry") mergers
which are important for the late-time growth of large passive systems.
Hence much work remains to be done in order to comprehensively understand
how galaxies evolve from primordial fluctuations into their present state.
It is hoped that the equilibrium model provides, in its simplicity,
a useful guiding framework for understanding the increasingly complex
problem of galaxy evolution.

 \section*{Acknowledgements}
The authors thank N. Bouch\'e, A. Dekel, R. Genzel, N. Katz, D.  Kere\v{s}, A.
Loeb, N. Murray, C. Papovich, E. Quataert, J. Schaye, L. Tacconi,
and D.  Weinberg for helpful discussions.  This work was supported
by the National Science Foundation under grant numbers AST-0847667
and AST-0907998.  Computing resources were obtained through grant
number DMS-0619881 from the National Science Foundation.

\end{document}